\documentclass{ws-procs9x6}

\begin{document}

\title{Explaining Low Redshift Quasar Evolution}

\author{Adam Steed and David H. Weinberg}

\address{The Ohio State University \\
140 West 18th Ave, \\ 
Columbus, OH 43210, USA\\ 
E-mail: asteed,dhw@astronomy.ohio-state.edu}

\maketitle

\abstracts{
We have developed a flexible framework for constructing physical 
models of quasar evolution that can incorporate a wide variety of observational 
constraints, such as multi-wavelength quasar luminosity functions (QLFs), 
estimated masses and accretion rates of active black holes, space densities of 
quasar host galaxies, clustering measurements, and the mass function of black 
holes in the local universe.  In this brief contribution we focus on the 
observed decline in the QLF break luminosity at $z<2$, which can be explained 
either by a shift toward lower characteristic accretion rates at low $z$ or by 
preferential suppression of activity in higher mass black holes.
}

The central actor in our treatment of quasar evolution 
(Steed \& Weinberg [2004]) is the accretion probability distribution 
$p(\dot{m}|M,z)$, the probability that 
a black holeof mass $M$ at redshift $z$ is accreting mass at a rate $\dot{m}$
in Eddington units.  The key supporting players are the black hole mass function 
$n(M,z)$ and a physical model of accretion that predicts the radiativeefficiency 
for a given $\dot{m}$.    At a given redshift, $p(\dot{m}|M,z)$ and $n(M,z)$ 
together determine the quasar luminosity function, and $p(\dot{m}|M,z)$ also 
determines the accretion driven growth of the black hole 
population, and hence the evolution of $n(M,z)$.

Boyle et al. (2000) find that the observed QLF at $z<2$ is well described by a 
double power-law with a break luminosity, $L_{\rm{brk}}$, that declines toward 
low z.  Since black hole masses themselves cannot decrease, we find that 
reproducing this behavior requires either a shift in $p(\dot{m}|z)$ that 
increases the relative probability of low accretion rates or an evolving mass 
dependence of $p(\dot{m}|M,z)$ that preferentially shuts off accretion onto 
high mass black holes at low $z$.  With regard to $\Phi(L)$ alone, the two 
models are effectively degenerate.  However, if the first mechanism dominates, 
then the QLF changes character between $z=2$ and $z=0$, shifting from a sequence
of black hole mass toward a sequence of accretion rates. If the second method 
dominates, then the QLF remains a sequence of black hole mass at all redshifts, 
predicting that luminous AGN at low redshift consist primarily of low mass 
black holes with a narrow range of $\dot{m}$ values that produce high 
$L/L_{\rm{edd}}$.  For details see Steed \& Weinberg (2004).

\begin{figure}[ht]
\centerline{\epsfxsize=3.8in\epsfbox{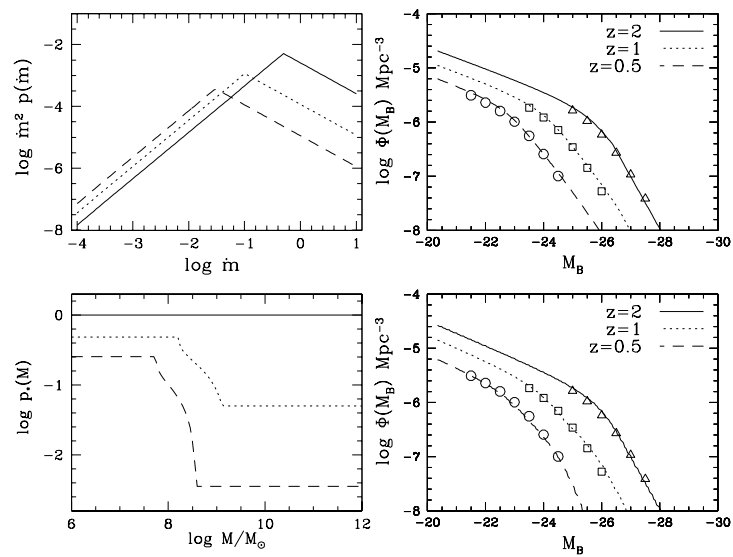}}   
\caption{Alternative explanations for Boyle et al.'s (2000) observed luminosity
evolution (points in the right hand panels).  Upper panels show a model in 
which $p(\dot{m},z)$ is independent of mass but evolves to lower characteristic 
$\dot{m}$.  Lower panels show a model in which the shape of $p(\dot{m})$ is 
fixed but the mass-dependent normalization $p_{*}(M)$ evolves, preferentially 
reducing the duty cycle of high mass black holes at low redshifts.  \label{evol}}
\end{figure}

\end{document}